\documentclass[11pt]{article}

\usepackage{eulervm}
\usepackage{setspace}
\usepackage[pdftex]{graphics}
\usepackage{graphicx}
\usepackage{color}
\usepackage{latexsym}
\usepackage{amsfonts}
\usepackage{amssymb, amsmath}
\usepackage{amsthm}
\usepackage{subcaption}
\usepackage{float}
\usepackage{bm}
\DeclareGraphicsExtensions{.pdf,.png,.jpg}

\usepackage{natbib} 
\usepackage{enumerate}

\usepackage[colorlinks=true, pdfstartview=FitV, linkcolor=blue, citecolor=blue, urlcolor=blue]{hyperref}

\usepackage{tikz}

\begin{document}
	
\title{A decomposition method to evaluate the\\ `paradox of progress' with evidence for Argentina \vspace{1cm}}

\author{Javier Alejo\footnote{IECON-Universidad de la Rep\'ublica, Uruguay. E-mail: javier.alejo@ccee.edu.uy} 
\and 
Leonardo Gasparini\footnote{CEDLAS-UNLP and CONICET, Argentina. e-mail: leonardo.gasparini@econo.unlp.edu.ar}
\and
Gabriel Montes-Rojas\footnote{Universidad de Buenos Aires and CONICET, Argentina. e-mail: gabriel.montes@fce.uba.ar}
\and
Walter Sosa-Escudero\footnote{UdeSA and CONICET,  Argentina. e-mail: wsosa@udesa.edu.ar}
}
	
\maketitle
	
\begin{abstract}
\noindent The `paradox of progress' is an empirical regularity that associates more education with
larger income inequality. Two driving and competing factors behind this phenomenon are
the convexity of the `Mincer equation' (that links wages and education) and the heterogeneity in its returns, as captured by quantile regressions. We propose a joint least-squares and quantile regression statistical framework to derive a decomposition in order to evaluate the relative contribution of each explanation. The estimators are based on the `functional derivative' approach. We apply the proposed decomposition strategy to the case of Argentina 1992 to 2015.
\vspace{7mm}
		
\vspace{3mm}
		
\end{abstract}
	
\onehalfspacing 
\newpage

\section{Introduction}

Intuitively, more education should be associated with less inequality, provided that access to education contributes to equal opportunities. Nevertheless,  the empirical evidence suggests otherwise: there is a positive association between average level of education and wage inequality, a phenomenon labeled as the `paradox of progress' by Bourguignon et al. (2004). Recent evidence on this effect can be found in Beccaria et al. (2015) and Ferreira et al. (2016), among others.

There are two alternative hypotheses that may help to rationalize this paradox. First, the `Mincer equation', that links (log) wages with its determinants, is found to be positive and convex with respect to education, hence higher levels of the latter are associated with higher wage inequality. Tinbergen (1972) and Sattinger (1993) are seminal references that may help explain why this convexity occurs, in a labor market that has a differential rent structure, and education serves as a signal to match workers to better jobs. In turn, this depends on the relative scarcity of capital stock across sectors. Thus the overall result is that a more equal distribution of education increases wage inequality due to the convexity of the returns to education (Legovini et al. 2005). Empirical papers that explore this line of research are Bourgignon, Ferreira and Lustig (2005)  and Battiston, García Domench and Gasparini (2014), among others.
Second, and independently of convexity, the paradox of progress may arise due to individual heterogeneity in returns to schooling. Becker and Chiswick (1966) introduce the idea that human capital depends on individual unobservable characteristics, hence returns to education are heterogeneous. And when positively associated with the conditional distribution of earnings, relatively richer individuals get more from education, hence increased levels of education are associated with a higher mean wage and more inequality. Empirical contributions on this line of research are based on quantile regression (QR), and include Buchinsky (1994, 2001), Martins and Pereira (2004), Staneva et al. (2010), Ariza and Montes-Rojas (2019), among others.

Each model implies a theoretical interpretation on the structure and behavior of the labor market, hence it is important to isolate the relative contribution of each of these two factors behind the increased levels of inequality associated with more education. This paper proposes an econometric framework that encompasses these two models and that leads to a natural way to quantify the absolute and relative contribution to inequality of convexity and heterogeneity. A functional framework is proposed using mean (i.e. ordinary least-squares, OLS) and QR models to estimate the relative contribution of each hypothesis. The proposed method uses functional derivatives as in Firpo et al. (2009) applied to the variance of the logarithms as the indicator of inequality.

The method is implemented for the case of Argentina, which offers a relevant empirical
case due to the large changes in wage inequality observed in last 30 years. The
obtained results show that at the beginning of the 1990's both unequalizing aspects had the
same relevance on the wage distribution. However, convexity of the mean returns became more relevant in 1998 and grows gradually in the following decade and a half, when the effect of
heterogeneity became statistically irrelevant around 2015.

This paper is organized as follows. Section \ref{Mincer} summarizes the main approaches to the paradox of progress. Section \ref{methodology} shows the econometric methodology. Section \ref{results} applies this method to Argentina 1992-2015. Finally, Section \ref{conclusion} concludes. Mathematical proofs are gathered in the Appendix.

\section{Mincer equation models}\label{Mincer}

The Mincer equation (Mincer, 1974) is a widely used hedonic price model that postulates that what the market pays for a good or factor depends on its observable characteristics. In the
labor market, wages depend on time invested to acquire knowledge to apply tasks that require
different degrees of complexity. The `paradox of progress' relates to the empirical finding that higher levels of education are related with higher mean wages and, counter intuitively, with more inequality. As advanced in the Introduction, there are two driving forces behind this phenomenon.

\subsection{Convexity of the returns to education}

\begin{figure}
 \caption{Mincer equations - Distributional effects of education}
 \label{fig:figura1.1}
 \centering
  \begin{tabular}{cc}
  (a) Convexity & (b) Heterogeneity\\
\includegraphics[width=0.5\textwidth]{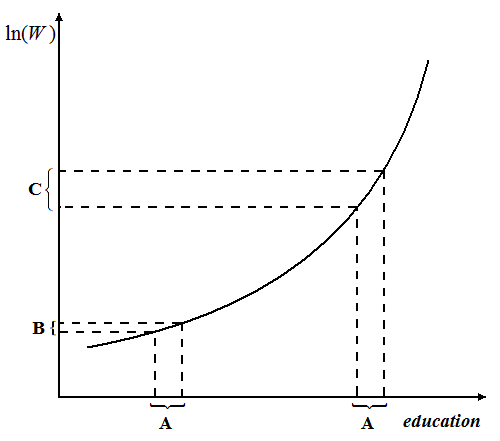}
&    \includegraphics[width=0.5\textwidth]{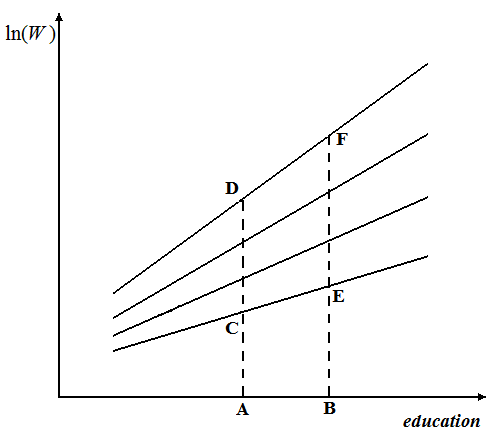}
\end{tabular}
\end{figure}

The first hypothesis that rationalizes more education with higher inequality states that, in
a partial equilibrium framework, the convexity of the returns to education in the Mincer
equation is the cause of the unequalizing effect of increasing the level of education. Figure \ref{fig:figura1.1} (a) illustrates this idea. A given increment in education ($A$) generates a higher effect for higher levels of education ($C$) than for lower levels of education ($B$). The convexity of the function implies that $C > B$. A formal proof of the result that more education leads to more unconditional inequality under convexity is presented in Appendix A.1.

A seminal theoretical explanation is given by Satinger (1993). The curvature of the wage
curve depends on a labor market where there are differential rents and workers self-select
into different occupations according to their skills. Years of schooling operate as a signalling mechanism about those skills. Firms have heterogeneous capital stocks and demand those skills. The matching is done by assuming that more capital is associated with higher skills requirements. In equilibrium, there is a positive association between wages and education and the functional form between these depends on the distribution of skills and capital stock. If capital has more dispersion than skills, then there is relative scarcity of high skilled workers. As a result, the market pays them relatively more vis-\`a-vis low skilled workers. The same model allows for the possibility of an increasing but concave function.

The convexity of Mincer equations with respect to education is a common empirical feature. The procedure to study this effect is by simulating (mean based) Mincer equations
estimated with household surveys, assigning one additional year of education uniformly across the entire population. These models typically assume homogeneous (and parametric) returns to schooling in a non-linear fashion to allow for convexity, and then the mean effects to the entire population. Bourgignon, Ferreira and Lustig (2005) and Battiston, García Domench and Gasparini (2014) are examples of this line of research, whose results provide strong evidence of the convex relation between (log) earnings and education, that may help explain the paradox of progress.

\subsection{Heterogeneity in returns to education}

The second line of research focuses on the role of unobservable factors interacting with education in the determination of wages. Becker and Chiswick (1966) present a model where
each individual has her own human capital, and returns to education are individual specific.
Thus wage dispersion corresponds to differences in unobservable factors (like, for example, intelligence or family background) conditional on the same observable variables. These unobservables were latter studied in several dimensions, including cognitive and non-cognitive skills. In turn, these may also be correlated with observable characteristics. That is, more education also corresponds to a higher ranking in unobservables. When these
factors complement with education in the determination of wages, higher level of education
lead to higher returns for individuals up in the conditional distribution of wages. For example, for a given level of education, individuals with better unobserved family background benefit more from education than those with worse conditions. Concretely, in such setup, education has a positive in impact on the mean wage and also in its conditional variance which, in turns, increases the (unconditional) wage inequality.

Mean based models do not capture these distributional effects of an increment in education.
Consider the example of Figure \ref{fig:figura1.1} (b). There are four different wage curves, one for each of
four specific quantiles of the conditional distribution of wages. The fact that the slopes of these curves are increasing illustrates the idea that individuals up in the conditional distribution (for example, with better family background) face higher returns to education. The wage gap for individual with education in $A$ is $D -C$,  smaller than that of individual with education $B$, $F -E$. Hence, more education leads to more conditional `within'
inequality, that eventually translates into more unconditional wage dispersion. A formal proof that more education leads to higher unconditional inequality under heterogeneity is shown in Appendix A.1. 

An extensive line of research indexes unobservable characteristics using the quantiles of the conditional distribution. Buchinsky (1994, 2001), Martins and Pereira (2004), Staneva et al. (2010), among many others, provide
strong evidence of this empirical pattern, where wage returns are increasing along quantiles and the wage gap is increasing in education.  Different econometric analyses use micro simulations after the estimation of QR models, as in Autor, Katz and Kearney (2005), Machado and Mata (2005), Melly (2005), Montes-Rojas, Siga and Mainali (2017), among others. 

An alternative approach is to handle unconditional effects directly, as in the recentered influence function (RIF) regression method of Firpo, Fortin and Lemieux (2009,2018), that leads to `unconditional' quantile regressions.

In any case, in their current state, mean models focus on measuring the contribution of
convexity, and their QR counterpart on that of heterogeneity. In the next section we propose a modeling strategy that encompasses both ideas and leads to a natural way to quantify the relative importance of each of the factors behind the paradox of progress.


\section{A decomposition approach for the paradox of progress}\label{methodology}

In this section we present a methodology to decompose the two potential effects driving the paradox of progress. It is based on a simple re-parameterization of a QR model using cross-sectional data, based on  Autor, Katz and Kearney (2005) model. We further assume the exogeneity of the covariates (and in particular schooling). The proposed method can be used with instrumental variables if available to tackle endogeneity as well.

Following the empirical literature we consider only partial equilibrium effects, which means that changes in the composition of education does not affect factor prices. This is the same framework of Firpo et al. (2009) where the conditional distribution can be used together with changes in covariates to study the unconditional wage distribution. 

\subsection{Population model}

Consider that wages are represented by the following  a random coefficient representation of QR (Koenker and Xiao (2006), Montes-Rojas et al. (2017))
\begin{equation}\label{eq:W}
 W = X' \alpha(U),
\end{equation}
where $W$ is the logarithm of wages, $X$ is a set of observable attributes  (education, experience, gender, etc.) and $U|X$ is a random variable with the standard uniform distribution representing the effect of unobservable components. Index $U$ represents the ranking of individuals along the distribution of wages conditional on $x$. \footnote{Using  $U$ as a uniform random variable is an application of the inverse transformation method to represent random variables by their quantiles. It requires that $F(w|x)$ is a continuous function (Devroye, 1986).} This ranking is determined by the unobservable components that explain wage heterogeneity (ability, luck, etc.). Therefore, the functional form of $\alpha(.)$ depends on their distribution . In turn, these coefficients can be interpreted as the interaction of observables and unobservables.

Following Autor, Katz and Kearney (2005), let $E(W|X=x)=x'\beta$ be the conditional mean model. Then we can write \eqref{eq:W} as

\begin{equation}\label{eq:W2}
 W=X' \beta+X' \gamma(U),	
\end{equation}
defining $\gamma(U) := \alpha(U) - \beta$. In other words,  $\gamma(U)$ is the difference between the $U$-quantile and the mean. In this setup, discrepancies in $W$ are due to \emph{between} differences, associate to different levels of $X$, and to \emph{within} differences, that is, wage disparities associated to different wages for the same level of $X$. 

This model will be rewritten in a form that contains the two explanations outlined above for the paradox of progress. First, following Buchinsky (1994) the heterogeneity in wages can be analyzed in terms of how $U$ affects wages through $\alpha$ or $\gamma$. Second, following Bourgignon, Ferreira and Lustig (2005) heterogeneity in wages could arise because of a non-linear (convex) functional form in $X$. For this we consider a simple model with $X=[H,H^2,Z]$, where $H$ is a continuous measure of human capital (i.e. years of schooling) and $Z$ other covariates.

The relevant measure of wage inequality ($I$) will be the variance of logs. In general, this measure correlates positively with other measures of inequality (e.g. Gini, Theil) but it allows for a simple regression framework. Using the law of total variance (see Angrist and Pischke (2009, p.33)):

\begin{equation}
 I := Var(W) = Var[E(W|X)]+E[Var(W|X)].	
\end{equation}
Then, calculating  $E(W|X)$ and $Var(W|X)$ for our model (see Appendix A.2 for details) and replacing we obtain the following wage decomposition:

\begin{equation}\label{eq:I}
 I = \beta' V \beta + tr(\Omega V) + E' \Omega E,		
\end{equation}
where  $V$ is the variance-covariance matrix of $X$, $E$ is the vector of (unconditional) means of $X$, and  $\Omega=Var[\gamma(U)]$ is a measure of the distance of the mean and quantile coefficients. 

\subsection{Decomposing the marginal effect of education}

Given that \eqref{eq:I} contains both the mean ($E$) and the variance ($V$) of $X$, the unconditional inequality $I$ not only depends on parameter heterogeneity but also on how covariates are distributed. Therefore, to study the marginal effect on the variance of (log) wages how covariates are shifted. Following Fortin et al. (2009,2018) we assume a small translation (location shift) in $H$, years of schooling, to $H+\epsilon$ with $\epsilon\to0$. 
Let $\delta(T(X)) :=\lim_{\varepsilon \to 0} \frac{T[(X+\varepsilon)] - T(X)}{\varepsilon}$ denote the marginal effect of this location shift on any statistic $T(.)$ due to a location shift in covariate $X$. 

Assuming that the parameters $\beta$ and $\gamma$ are not affected by this shift, then the total effect of a location shift in $X$ on inequality can be expressed as follows (see Appendix A.3 for a derivation)

\begin{equation}\label{eq:I2}
  \delta(I) = \beta' \delta(V) \beta + tr[\Omega \delta(V)] + 2E' \Omega \delta(E).
\end{equation}

Consequently, changes can be decomposed into a \textit{between} and a \textit{within} effect as follows:

\begin{align*}
  EF_{betw} &= \beta' \delta(V) \beta  \\
  EF_{with}  &= tr[\Omega \delta(V)] + 2E' \Omega \delta(E)
\end{align*}
such that  $ \delta(I)=EF_{betw}+EF_{with}$. On one hand, education affects inequality by affecting the average wage gap among different groups. On the other hand, there is an additional effect coming from the within group inequality due to differences across quantiles. 

Consider a simple example to study these effects. For simplicity we assume that $X=[H,H^2]$. Then, 
\begin{equation*}
  W = \beta_0 + \beta_1 H + \beta_2 H^2 + \gamma_0 (U) + \gamma_1 (U)H + \gamma_2 (U) H^2.
\end{equation*}
Then, applying the decomposition discussed above:

\begin{align*}
  EF_{betw} &= 4(\beta_1 V_{11} + \beta_2 V_{12} ) \beta_2,  \\
  EF_{with}  &= 2[\Omega_{01}+2\Omega_{02} E_1+3\Omega_{12} E_2+\Omega_{11} E_1+2\Omega_{22} E_3],
\end{align*}
where $V_{ij}$, $\Omega_{ij}$ and $E_i$ correspond to the $i,j=0,1,2$ or $i=0,1,2$ elements in the matrix or vector. Convexity is led by $\beta_2 \neq 0$. Note that even though convexity is the leading factor behind the  the between inequality it does not necessarily play a role in the within effect. Hence, education might increase inequality through the within channel even when the effect of education is linear in the `mean' part of the model ($\beta_2 = 0$) . 

Consider the following three illustrative cases:

\begin{itemize}

\item \textbf{Case 1 (linear homoskedastic model)}: $\beta_2=0$ and $\Omega$ has zeros except for $\Omega_{00}>0$. Then, $EF_{betw}=EF_{with}=0$.

\item \textbf{Case 2 (quadratic homoskedastic model)}: $\beta_2>0$ and $\Omega$ has zeros except for $\Omega_{00}>0$. Then,

\begin{align*}
  EF_{betw} &= 4 V_{12} \beta_2 ^2 > 0,  \\
  EF_{with}  &= 0.
\end{align*}

\item \textbf{Case 3 (linear heteroskedastic model)}: $\beta_2=0$ and the diagonal elements of $\Omega$ are non zero.\footnote{We could also consider the non-diagonal elements for a more involved model.} Then,

\begin{align*}
  EF_{betw} &= 0,  \\
  EF_{with}  &= 2\Omega_{11} E_1+4\Omega_{22} E_3>0.
\end{align*}

\end{itemize}

This decomposition allows us to consider the separate effects of the two potential explanations for the paradox of progress. In turn, this should be evaluated on empirical terms.

\subsection{Estimation and inference}
The proposed decomposition leads to a simple way to quantify and separate the relative contribution of the factors behind the paradox of progress. This subsection describes how to
implement it in practice with a sample of wages and its determinants. Let $\{w_i,x_i\}_{i=1}^n$ be a random sample of waged employees, where $x$ contains $h$ (education) and $h^2$ and other individual characteristics $z$. The parameters $\beta$ and $\alpha(\tau)$, $\tau \in (0,1)$, are estimated by OLS and QR as in Koenker and Basset (1978) and Koenker (2005).

In order to estimate $\Omega$, the variance-covariance matrix of $\gamma(U)=\alpha(U)-\beta$ where $U$ is uniformly distributed on $[0,1]$, we use the following procedure. Consider a grid of $M$ indexes, $0<\tau_1<\tau_2<...<\tau_M<1$, and let $\alpha(\tau_m)$ be the corresponding QR coefficient. Then,
\begin{equation*}
	\hat{\Omega}=M^{-1} \sum_{m=1}^{M} [\hat{\alpha}(\tau_m)-\hat{\beta}]\cdot[\hat{\alpha}(\tau_m)-\hat{\beta}]'	
\end{equation*}
If $M$ is large enough, which in turn should depend on $n$, i.e. $M=M_n$, then $\hat{\Omega}\stackrel{p}{\rightarrow}\Omega$ as $n\rightarrow\infty$. Here we propose to invoke a similar procedure to Chernozhukov, Fernández-Val and Melly (2013), where they impose that $t=\tau_1$ and $\tau_M=1-t$ for a small $t>0$ (to avoid estimating extreme quantiles) and the quantiles are estimated on a fine mesh with mesh width $\psi=\tau_j-\tau_{j-1},\ j=2,...,M$, with $\psi\sqrt{n}\rightarrow0$ as $n\rightarrow\infty$.


Finally, consider the estimators of $E$ and $V$. Let $Q$ be the number of regressors included in $z$, then we obtain $\delta(E)$ and $\delta(V)$ (see Appendix A.4 for a derivation):

\begin{align}\label{eq:I3}
\delta(E) =\left[
\begin{matrix}
0 \\
1 \\
2E_1 \\
0_{1 \times Q} 
\end{matrix} \right]
& \and & 
\mbox{~and~} 
& \and & 
\delta(V) =\left[
\begin{matrix}
0 & 0 & 0 & 0_{1 \times Q}\\
0 & 0 & V_{11} & 0_{1 \times Q}\\
0 & V_{11} & 2V_{12} & M_{1z}\\
0_{Q \times 1} & 0_{Q \times 1} & M_{z1} & 0_{Q \times Q}
\end{matrix} \right]
\end{align}
where $E_1=E(h)$, $V_{12}=Cov(h,h^2 )$, $M_{z1}'={1z}=Cov(h,z)$ ($1\times Q$ vector) and $0_{A \times B}$ is a null matrix of dimensions $A \times B$. All of these components can be estimated. We use
\begin{align*}
  \hat{E} &= n^{-1} \sum_{i=1}^{n} x_i  \\
  \hat{V} &= (n-1)^{-1} \sum_{i=1}^{n} (x_i - \hat{E}) \cdot (x_i - \hat{E})'.
\end{align*}
Then, replacing in \eqref{eq:I2},

\begin{equation}\label{eq:estI}
  \hat{\delta}(I) = \hat{\beta}' \delta(\hat{V}) \hat{\beta} + tr[\hat{\Omega} \delta(\hat{V})] + 2\hat{E}' \hat{\Omega} \delta(\hat{E}).
\end{equation}

The first term in \eqref{eq:estI} is the estimation of the convexity effect in returns to education, and the last two are due to the heterogeneity in returns to human capital. Thus this method allows us to disentangle the two effects.

Bera, Galvao and Wang (2014) study the asymptotic properties of the joint estimators of the OLS and QR estimators, and derive uniform weak convergence of the joint process indexed by the quantile index, which justifies the application of bootstrap procedures. Montes-Rojas, Siga and Mainali (2017) use this strategy to study parameter heterogeneity comparing mean and quantile coefficients. For the purposes of this paper, the proposed measure of inequality and its components are continuous transformations of estimators of OLS and QR coefficients, hence the continuous mapping theorem applied to the components of \eqref{eq:estI} guarantee the existence of a stable asymptotic distribution and justifies the use of the wild bootstrap.

\section{Exploring the paradox of progress for the case of Argentina}\label{results}

In this section we implement the proposed decomposition methodology with data from the
Permanent Household Survey (EPH, acronym in Spanish) implemented by the National Institute of Statistics and Censuses (INDEC) of Argentina. The decomposition is computed in four
distant years to explore different moments of the Argentine wage distribution and to evaluate
long-run changes: 1992, 1998, 2008 and 2015. The criterion for choosing these years was due
to data availability and to take periods with a relative macroeconomic stability and similar
survey and sampling methodologies. The first two years belong to the so-called discontinuous
survey methodology, usually carried out in the months of May and October, while the last two
years correspond to the continuous survey methodology, carried out quarterly. INDEC has
been expanding the sampling coverage over the period of analysis, and therefore we only use
the observations that belong to the urban agglomerates present in the 1992 EPH to keep all
the estimates comparable.3 The data correspond to the surveys collected during the second
semester with the exception of 2015, which is when the EPH changed its survey methodology
in the second half of the year and therefore we used the first semester for comparability reasons. The sample used in all cases is of men between 16 and 65 years of age.

Table \ref{table4.1} presents the estimated coefficients for the education variables of the Mincer equations of the conditional mean and some relevant conditional quantiles. These regressions are
only descriptive, for the decomposition a much larger number of quantiles is used. Other
typical covariates from the Mincer equations literature are also included. Quadratic term are statistically significant in the four years considered and at the different points of the conditional distribution. This means that the relationship between wages and educational level is  convex.From the perspective of the theoretical models in Sattinger (1993), this is interpreted as an indication of a certain relative scarcity of human capital (measured with years of education) in relation to physical capital.
 
\begin{figure}
\caption{Convexity and Heterogeneity of the returns to education.}
\label{fig:figura4.1}
\centering
\includegraphics[scale=0.12]{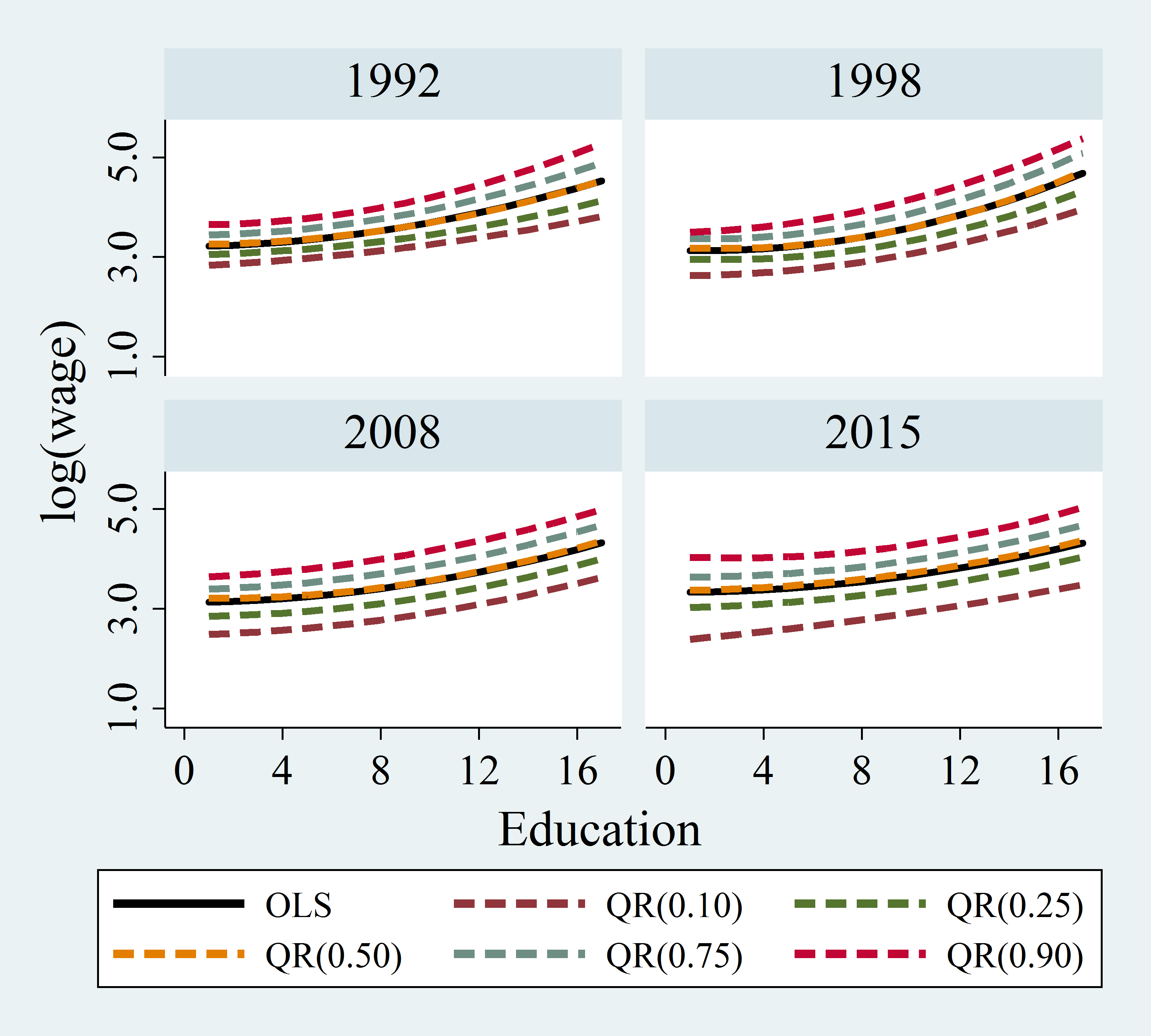}

Source: own estimates based on the EPH.

Note: the other covariates are evaluated at their sample means.
\end{figure}

Additionally, the coefficients of the conditional quantiles are not constant, suggesting a
heterogeneous pattern in the returns to education, similar to that postulated in Becker and
Chiswick (1966). For example, comparing the coefficient of the quadratic term between conditional deciles, the difference is high in 1992 and 2015 (the last decile is at least 2.5 times the first), slight in 1998 (1.3 times) and null in 2008. Figure \ref{fig:figura4.1} shows this result graphically comparing the predictions of each Mincer regression equation (keeping the rest of the covariates in their sample mean). Convexity appears to be more relevant in 1998, while the pattern of heterogeneity in returns to education has been disappearing over the last fifteen years.

\begin{table}
	\caption{Partial relationship between wage (log) and educational level. Argentina 1992 - 2015.}\label{table4.1}
	\centering
\begin{tabular}{rrrrrrr}
\hline
    {\bf } &  {\bf OLS} & {\bf QR(0.10)} & {\bf QR(0.25)} & {\bf QR(0.50)} & {\bf QR(0.75)} & {\bf QR(0.90)} \\
\hline
           &            &            &            &            &            &            \\

           &                                 \multicolumn{ 6}{c}{{\bf 1992 ($n = 12196$)}} \\

Education &     0.0070 &  0.0228*** &  0.0062*** & -0.0022*** &     0.0006 & -0.0049*** \\

           &     (1.00) &    (26.08) &     (8.89) &    (-5.05) &     (0.80) &    (-4.95) \\

Education Squared &  0.0042*** &  0.0021*** &  0.0034*** &  0.0045*** &  0.0050*** &  0.0059*** \\

           &    (12.73) &    (52.03) &   (103.12) &   (219.24) &   (152.62) &   (127.97) \\

           &            &            &            &            &            &            \\

           &                                 \multicolumn{ 6}{c}{{\bf 1998 ($n = 11228$)}} \\

Education & -0.0198*** & -0.0056*** & -0.0289*** & -0.0334*** & -0.0240*** &  0.0035*** \\

           &    (-2.76) &    (-4.94) &   (-52.82) &   (-91.66) &   (-70.88) &     (6.04) \\

Education Squared &  0.0065*** &  0.0049*** &  0.0063*** &  0.0072*** &  0.0073*** &  0.0063*** \\

           &    (19.58) &    (93.00) &   (249.73) &   (423.64) &   (464.61) &   (237.31) \\

           &            &            &            &            &            &            \\

           &                                 \multicolumn{ 6}{c}{{\bf 2008 ($n = 14580$)}} \\

Education &     0.0034 &  0.0103*** &     0.0004 & -0.0103*** &  0.0071*** &  0.0168*** \\

           &     (0.51) &    (14.10) &     (0.83) &   (-28.85) &    (15.52) &    (25.14) \\

Education Squared  &  0.0039*** &  0.0034*** &  0.0040*** &  0.0046*** &  0.0041*** &  0.0037*** \\

           &    (13.12) &   (104.41) &   (209.74) &   (286.76) &   (198.73) &   (125.35) \\

           &            &            &            &            &            &            \\

           &                                \multicolumn{ 6}{c}{{\bf 2015 ($n = 14553$)}} \\

Education &    -0.0010 &  0.0432*** &  0.0063*** &  0.0018*** & -0.0065*** & -0.0279*** \\

           &    (-0.14) &    (41.27) &    (10.41) &     (4.53) &   (-12.42) &   (-57.34) \\

Education Squared &  0.0035*** &  0.0014*** &  0.0032*** &  0.0034*** &  0.0040*** &  0.0050*** \\

           &    (11.28) &    (29.98) &   (118.39) &   (192.49) &   (171.48) &   (233.17) \\

           &            &            &            &            &            &            \\
\hline
\end{tabular}  

Notes: Other variables included are: potential experience (and its square), marital status, and controls by geographic region. The t statistics are shown in parentheses, * indicates significance at 10\%, ** at 5\% and *** at 1\%.

\end{table}

The presence of heterogeneities and a convex relationship between (log) wages and education indicates that increased education is associated with higher unconditional inequality, as discussed in Section 2. In order to quantify the strength of these effect, and to decompose their relative importance in increase inequality, Table 2 presents the results of the decomposition methodology introduced in this paper. The first two rows of the Table shows the evolution of wage inequality for men aged 16 to 65, measured with the Gini index and the variance of the logarithms. Both indicators show a similar evolution: an increase in wage inequality towards the end of the 1990s, a relevant distributional improvement in the 2000s, slightly sustained towards the middle of the last decade.

As we discussed in Section \ref{Mincer}, both the degree of convexity and heterogeneity in the Mincer equation are two key factors behind the partial effect of education on wage inequality. Regressions such as those in Table \ref{table4.1} are analogous to those used in the literature that analyzes the effect of education on wage inequality. However, they are not enough to quantify the change in the unconditional distribution or measure the relative weight of convexity and heterogeneity as explanatory factors. The contribution of this paper is to use the decomposition methodology proposed in Section 3 for this purpose.

\begin{table}
	\caption{Marginal effect of education on inequality.}\label{table4.2}
	\centering
\begin{tabular}{rrrrr}
\hline
    {\bf } & {\bf 1992} & {\bf 1998} & {\bf 2008} & {\bf 2015} \\
\hline
           &            &            &            &            \\

1. Inequality &            &            &            &            \\

{\it       Gini index} &       40.5 &       44.0 &       39.8 &       35.6 \\

{\it       Variance of logarithms} &       42.5 &       53.8 &       48.5 &       42.1 \\

           &            &            &            &            \\

2. Numerical simulation (location shift) &     1.82** &     3.69** &     1.71** &     1.17** \\

           &     (0.33) &     (0.39) &     (0.24) &     (0.19) \\

3. RIF estimate (Firpo et al. 2009) &     4.65** &     6.36** &     2.03** &     1.36** \\

           &     (0.51) &     (0.56) &     (0.39) &     (0.38) \\

           &            &            &            &            \\

4. Quantile decomposition (levels) &            &            &            &            \\

       Between effect (convexity) &     1.81** &     3.69** &     1.71** &     1.17** \\

    {\it } &     (0.33) &     (0.39) &     (0.23) &     (0.19) \\

       Within  effect (heterogeneity) &     1.88** &     1.47** &      0.56* &      0.540 \\

    {\it } &     (0.28) &     (0.32) &     (0.28) &     (0.30) \\

      Total change &     3.69** &     5.16** &     2.27** &     1.71** \\

           &     (0.48) &     (0.55) &     (0.35) &     (0.34) \\

           &            &            &            &            \\

5. Quantile decomposition (percentage) &            &            &            &            \\

      Between effect (convexity) &     49.1\% &     71.5\% &     75.3\% &     68.4\% \\

      Within effect (heterogeneity) &     50.9\% &     28.5\% &     24.7\% &     31.6\% \\

       Total change &      100\% &      100\% &      100\% &      100\% \\
\hline
\end{tabular}  

Notes: Other variables included are: potential experience (and its square), marital status, and controls by geographic region. The t statistics are shown in parentheses, * indicates significance at 10\%, ** at 5\% and *** at 1\%.

\end{table}

The second block of Table \ref{table4.2} presents a numerical simulation exercise similar to those in the literature based on the conditional mean. It uses the Mincer equation estimated by OLS and its residuals to re-compute a new wage distribution after a small horizontal translation of the education covariate. Specifically, if an individual $i$ has an education level of $h_i$, the exercise is to impute $h_i+\varepsilon$ years of education and construct a counterfactual salary (in logarithms) $w_i^s$ through the prediction of the conditional mean and adding the OLS residual. Then, the marginal variation on inequality is calculated as $[V(w^s )-V(w)]/\varepsilon$, where $\varepsilon$ is a small value. \footnote{The value used in this paper is $\varepsilon=0.01$.} The next block of Table \ref{table4.2} contains the change in variance estimated with the Firpo et al. (2009) methodology through RIF regression.\footnote{Although not reported, similar qualitative results are obtained if using the RIF for Gini.} Both methodologies show an unequalizing effect of education on wages, reaching its maximum impact in the late 1990s. Although both methodologies measure the effects of a small location shift on education, the gap between both is that the RIF regression is more accurate because it contemplates the entire information of the conditional distribution of wages, while the numerical simulation only extrapolates the behavior of the conditional mean. However, it is not possible to split the contributions of non-linearity and heterogeneity of the Mincer equation from the RIF regression. In this aspect, our decomposition method is really useful.

The fourth and fifth blocks of Table \ref{table4.2} show the results of implementing our proposed decomposition exercise. We set a grid of values for $\tau={0.005,0.01,…, 0.99,0.995}$ for the QR estimation, involving a total of $M=199$ estimated equations for each year.\footnote{This number of quantiles is similar to Melly (2005) for generating counterfactual distributions with many QR models.} The results show that the total effect is quite similar to the estimate made by the RIF method. The contribution of our decomposition is that it explicitly shows the relative weight of the empirical explanations of the unequalizing effect of education based on both the conditional mean and the conditional quantiles. This last aspect is somewhat arcane in RIF regression methodologies.

Almost all terms in the decomposition terms are statistically significant at the usual levels. At the beginning of the 90s convexity and heterogeneity had the same relevance on the increased inequality due to more education. However, the effect of convexity on the mean returns becomes more relevant in 1998 (just over 70\%) and seems to grow gradually in the following decade and a half, where the effect of heterogeneity becomes statistically irrelevant in 2015.

Using the theoretical framework discussed in Section \ref{Mincer}, this gap in the relevance of the two unequalizing forces could give rise to some interpretations of how the functioning of the labor market has changed. The increase in the relevance of convexity in the returns to education could be indicating a certain relative scarcity of supply in the stock of available human capital in relation to the variety of the type of qualified tasks demanded by employers. Consequently, the market pays more than proportionally more educated workers, who are likely to be able to work multiple tasks. On the other hand, the decreasing role of the heterogeneity in returns to education could be associated with a less important role of the market's unobservable wage determinants (innate skills, tenacity, intelligence, luck, etc.). This would indicate some direct loss in the return received by these unobservable factors in the market, or a decreased complementarity with education.

\section{Concluding remarks}\label{conclusion}

This paper proposes a decomposition methodology for the marginal effect of education on unconditional wage inequality. The method quantifies the relevance of two empirical arguments previously outlined in the literature to explain the unequalizing effect of education, labeled as the `paradox of progress'. The first one is based on the convexity of the Mincer equation interpreted as a conditional mean, while the other focuses on the heterogeneity of returns to education at different levels of the conditional distribution of (log) wages.

The key idea of our decomposition is based on a `functional derivative' as in in Firpo et al. (2009) applied to the variance of the logarithms as the indicator of inequality. Our proposal only requires consistent estimates of the parameters of the conditional mean and quantiles. The method applies naturally to any consistent regression (instrumental variables, panel data, etc.).

The implementation of the decomposition for the case of Argentine shows that in the early
1990s both convexity and heterogeneity had the same importance as unequalizing factors due
to increased education. Still, towards the end of the period studied, the effect of convexity dominates. This change in the relevance of the two unequalizing forces could give rise to some interpretations of how the functioning of the labor market has changed. On the one hand, the increase in the relevance of convexity in the returns to education could be indicating a certain relative scarcity of supply in the stock of available human capital in relation to the variety of the type of qualified tasks demanded by employers. Consequently, the market pays more than proportionally more educated workers, who are likely to be able to work on multiple tasks. On the other hand, the decreasing role of the conditional heterogeneity in returns to education would be associated with a less important role of the market's unobservable wage determinants (innate skills, tenacity, intelligence, luck, etc.). This would indicate some direct loss in the return received by these unobservable factors in the market, or indirectly due to a lesser complementarity of these with the educational level of the individuals. A detailed study behind these mechanisms is a relevant topic of further research.

\pagebreak

\section*{References}

\noindent Ariza, J. and Montes-Rojas, G. (2019). “Decomposition methods for analyzing inequality changes in Latin America 2002-2014''. \textit{Empirical Economics} 57(6), 2043-2078. 

\vspace{0.4cm}

\noindent Autor, D., Katz, L. and Kearney, M. (2005). "Rising wage inequality: The role of composition and prices''. NBER Working Paper No. 11628.

\vspace{0.4cm}


\noindent Battiston, D., García Domench, C. and Gasparini, L. (2014). ``Could an Increase in Education  Raise Income Inequality? Evidence for Latin America''. \textit{Latin American Journal of Economics} 51(1), 1-39.

\vspace{0.4cm}

\noindent Beccaria L., Maurizio R. and Vázquez G. (2015). ``Recent decline in wage inequality and formalization of the labour market in Argentina''. \textit{International Review of Applied Economics} 29(5), 677–700

\vspace{0.4cm}

\noindent Becker, G. and Chiswick, B. (1966). ``Education and the distribution of earnings''. \textit{American Economic Review} 56(1/2),  358-369. 

\vspace{0.4cm}

\noindent Bera, A., Galvao, A. Wang, L. (2014). ``On testing the equality of mean and quantile effects''. \textit{Journal of Econometric Methods} 3(1), 47-62.

\vspace{0.4cm}

\noindent Bourguignon, F., Lustig, N. and Ferreira. F. (2004). \textit{The Microeconomics of Income Distribution Dynamics}. Oxford University Press, Washington.

\vspace{0.4cm}

\noindent Buchinsky, M. (1994). ``Changes in the U.S. wage structure 1963-1987: Application of quantile regression''. \textit{Econometrica} 62(2), 405-458.

\vspace{0.4cm}

\noindent Buchinsky, M. (2001). ``Quantile regression with sample selection: Estimating women’s return to education in the U.S.''. \textit{Empirical Economics} 26, 87-113.

\vspace{0.4cm}




\noindent Chernozhukov, V., Fernández-Val, I. and Melly, B. (2013). “Inference on counterfactual distributions''. \textit{Econometrica} 81 (6),  2205-2268.

\vspace{0.4cm}



\noindent Firpo, S., Fortin, N. and Lemieux, T. (2009). ``Unconditional quantile regressions''. \textit{Econometrica} 77(3), 953-973. 

\vspace{0.4cm}

\noindent Firpo, S., Fortin, N. and Lemieux, T. (2018). ``Decomposing wage distributions using recentered influence function regressions''. \textit{Econometrica} 6(3): 41.

\vspace{0.4cm}






\noindent Koenker, R. (2005). \textit{Quantile Regression}. Cambridge University Press, Cambridge. 

\vspace{0.4cm}


\noindent Legovini A., Bouilloon C. and Lustig N. (2005). ``Race and gender in the labor market.'' In: Bourguignon, F., Ferreira, F., Lustig, N. (eds) The Microeconomics of Income Distribution Dynamics, Chapter 8, 275-312. Oxford University Press, New York.

\vspace{0.4cm}

\noindent Martins, P.S. and Pereira, P.T. (2004). ``Does education reduce wage inequality? Quantile regressions evidence from 16 countries".  \textit{Labour Economics} 11(3), 355-371.

\vspace{0.4cm}

\noindent Mata, J. and J. Machado (2005). ``Counterfactual decomposition of changes in wage distributions using quantile regression''. \textit{Journal of Applied Econometrics} 20, 445-465.

\vspace{0.4cm}

\noindent Melly, B. (2005). "Decomposition of differences in distribution using quanitle regressions". \textit{Labour Economics} 12, 577-90.

\vspace{0.4cm}

\noindent Mincer, J. (1974). "Schooling, experience, and earnings". National Bureau of Economic Research, Inc. NBER Books Series.

\vspace{0.4cm}

\noindent Montes-Rojas, G., Siga, L. and  Mainali, M. (2017). ``Mean and quantile regression Oaxaca-Blinder decompositions with an application to caste discrimination”.  \textit{Journal of Income Inequality} 15(3), 245-255.

\vspace{0.4cm}


\noindent Sattinger, M. (1993). "Assignment models of the distribution of earnings". \textit{Journal of Economic Literature} 31(2), 831-880. 
\vspace{0.4cm}

\noindent See, C. and Chen, J. (2008). ``Inequalities on the variances of convex functions of random variables.'' \textit{Journal of Inequalities in Pure \& Applied Mathematics}, 9. 

\vspace{0.4cm}

\noindent Staneva, A., Arabsheibani, R. and Murphy, P. (2010). ``Returns to education in four transition countries: Quantile regression approach". IZA Discussion Papers 5210.

\vspace{0.4cm}

\noindent Tinbergen, J. (1972). ``The impact of education on income distribution". \textit{Review of Income and Wealth} 18(3), 255-265.

\vspace{0.4cm}

\newpage

\newpage
\section*{Appendix}

\subsection*{A.1 Convexity and heterogeneity lead to higher unconditional inequality}

In this Appendix we establish two results. The first one shows that a location shift in a convex mincer equation leads to more unconditional inequality. The second one, that a location shift in a linear quantile regression model with increasing heterogeneity leads to more unconditional inequality. We make use of two results in See and Chen (2008).

\vspace{1cm}

\noindent \emph{Convexity leads to higher inequality:} Let $Y = f(X+ \epsilon)$, where $X$ is a random variable and $f$ is a differentiable, increasing and  convex function. Then

\[\frac{ d\; V(Y)}{ d\; \epsilon} \geq 0\]

\noindent \emph{Proof:} Start from:

\begin{center}
    $V[f(X+\epsilon)]=E\{f^2(X+\epsilon)\}-E^2[f(X+\epsilon)]$
\end{center}

Taking derivatives with respect to $\epsilon$, and using Lemma 2.2 in See and Chen (2008):

\begin{center}
    $ \frac{d}{d \epsilon}  V[f(X+\epsilon)]=E\{2f(X+\epsilon)\frac{d}{d \epsilon}f(X+\epsilon)\}-2E[f(X+\epsilon)]E[\frac{d}{d \epsilon}f(X+\epsilon)]$
\end{center}

Let $h(X):=f(X+\epsilon)$ and $g(X) := \frac{d}{d \epsilon}f(X+\epsilon)$, then:

\begin{center}
    $ \frac{d}{d \epsilon}  V[f(X+\epsilon)] = 2E\{h(X)g(X)\}-2E[h(X)]E[g(X)] \geq 0 $ 
\end{center}
by Lemma 2.1 in See and Chen (2008), since both $h$ and $g$ are increasing under the assumptions about $f$.

\vspace{1cm}

\noindent \emph{Heterogeneity lead to more inequality:} Now assume $Y=f(X+\epsilon,U)$, where $U$ represents unobserved heterogeneity and $X$ is a random variable. Consider the case $f(X+\epsilon,U)=a(X+\epsilon)+b(X+\epsilon)\;g(U)$, with $a(X+\epsilon)=a_0 + a_1 (X+ \epsilon) $, where $b$ and $g$ are positive and increasing functions, and $U \sim U(0,1)$ and independent of $X$. Assume $E[g(U)]=0$  and $V[g(U)]=\sigma_{g}^{2}$. Then

\[\frac{ d\; V(Y)}{ d\; \epsilon} >0.\]

\noindent \emph{Proof:} Start from:
\begin{center}
    $V(Y) = V[E(Y|X)] + E[V(Y|X)]$ 
\end{center}
Note that $E(Y|X)=a_0 + a_1\;(X+\epsilon)$ and $V(Y|X)=b(X+\epsilon)^2\sigma_{g}^{2}$. Then:

\begin{center}
    $V(Y) = a_{1}^{2}\;V(X) + E[b^2(X+\epsilon)]\sigma_{g}^{2}$ 
\end{center}

Using Lemma 2.2 in Chuen-Teck y Chen (2008):

\begin{center}
    $ \frac{d}{d \epsilon}  V(Y') = E[2b(X)c(X)]\sigma_{g}^{2} =2E[b(X)c(X)]\sigma_{g}^{2} $ 
\end{center}

\noindent where $c(X):=\frac{\partial}{\partial \epsilon}b(X+\epsilon)$. The result follows since $b$ is positive and increasing.

\subsection*{A.2 Derivation of eq. \eqref{eq:I}}

Consider equation \eqref{eq:W} and calculate the expectation of $W$ conditional on $X=x$, then
\begin{equation}
E(W|x)=x' E[\alpha(U)|x]=x' E[\alpha(U)] := x'\beta. \tag{a.1}\label{eq:a.1}
\end{equation}
Note that $\beta$ has been defined as the expectation of the random vector $\alpha(U)$, where $U|x$ is $Uniform(0,1)$. That is, the parameters $\beta$ of the conditional expectation are the average of the parameters of all the conditional quantiles.

On the other hand, consider \eqref{eq:W2} and compute the conditional variance
\begin{equation}
  Var(W|x) =Var[x'\beta|x] + Var[x'\gamma(U)|x] = x' Var[\gamma(U)] x := x' \Omega x, \tag{a.2}\label{eq:a.2}
\end{equation}
where $\Omega$ has been defined as the matrix of variances of the vector $\gamma(U)$. Note that by construction the expectation $E[\gamma(U)]=E[\alpha(U)-\beta]=0$ and therefore $Var[\gamma(U)]=E[\gamma(U)\gamma(U)' ]=\Omega$. That is, the matrix $\Omega$ is a notion of distance between the mean and the quantiles of the distribution $W|X$.

Combining and using the Law of Iterated Variances,
\begin{equation*}
Var(W)=Var[X' \beta]+E[X' \Omega X]
\end{equation*}
Then, using properties of variance for the product of vectors and properties of expectation for quadratic forms:
\begin{equation*}
	Var(W) = \beta' Var(X)\beta + tr[\Omega Var(X)] + E(X)' \Omega E(X).
\end{equation*}

\subsection*{A.3 Derivation of eq. \eqref{eq:I2}}

\eqref{eq:I2} is a particular case of the notion of (partial) functional derivative proposed by Firpo et al. (2009). In this literature it is usual to assume that the distribution of $w|x$ is not affected by changes in the distribution of $x$. This assumption translated into our quantile model means that the parameters $\beta$ and $\Omega$ do not change as a consequence of a location shift in any of the regressors included in $x$. Intuitively, this assumption makes explicit the fact that it is a partial equilibrium analysis, in the sense that a small change in education (measured by $ h $) does not change the returns to education. The functional derivative of the inequality $I$ with respect to a horizontal translation in $h$ is obtained by computing a differential limit of equation (3.4). For example, to derive the expression $\beta'V\beta$ (first term of $I$) we solve the following limit:

\begin{align*}
  \delta[\beta' Var(x) \beta] 	&= \lim_{\varepsilon \to 0} \frac{[\beta' Var(x+\varepsilon)\beta] - [\beta' Var(x) \beta]}{\varepsilon} \\
  						&= \lim_{\varepsilon \to 0} \frac{\beta' [Var(x+\varepsilon) -  Var(x)] \beta}{\varepsilon} \\
  						&= \beta' \left[ \lim_{\varepsilon \to 0} \frac{Var(x+\varepsilon) -  Var(x)}{\varepsilon}\right] \beta := \beta' \delta(V) \beta
\end{align*}

Using the previous reasoning but applied to the rest of the terms of equation \eqref{eq:I}, it follows that:
\begin{align*}
\delta[tr(\Omega V)] = tr[\Omega \delta(V)]
& \and & 
\mbox{~and~} 
& \and & 
\delta(E' \Omega E) = 2E'\Omega \delta(E)
\end{align*}

Finally, adding these three components gives equation \eqref{eq:I2} as a result.

\subsection*{A.4 Derivation of eq. \eqref{eq:I3}}

To obtain the expressions \eqref{eq:I3} it is convenient to analyze each of the elements in $E$ and $V$. The matrix $V$ contains all the variances and covariances of the variables included in the vector $x$, while $E$ is a vector that contains the expectation $x$. Formally,

\begin{align*}
E :=\left[
\begin{matrix}
E_0 \\
E_1 \\
E_2 \\
E_z 
\end{matrix} \right]
& \and & 
\mbox{~and~} 
& \and & 
V :=\left[
\begin{matrix}
V_{00} & V_{01} & V_{02} & V_{0z}\\
V_{10} & V_{11} & V_{12} & V_{1z}\\
V_{20} & V_{21} & V_{22} & V_{2z}\\
V_{z0} & V_{z1} & V_{z2} & V_{zz}
\end{matrix} \right]
\end{align*}

where the element notation includes the following scalars,
\begin{align*}
E_k=E(h^k)
& \and & 
\mbox{~and~} 
& \and & 
V_{jk}=Cov(h^j,h^k )
\end{align*}

for $k=0,1,2$ y $j=0,1,2$, together with the following $(Q\times1)$ vectors
\begin{align*}
E_z=E(z)
& \and & 
\mbox{~and~} 
& \and & 
M_{kz}=Cov(h^k,z)=M_{zk}'
\end{align*}

for $k=0,1,2$, and the $(Q \times Q)$ matrix
\begin{align*}
M_{zz}=V(z)
\end{align*}

Note that when $k=0$ the vector $M_{0z}$ is a null vector, because it is the covariance between $h^0=1$ with each of the regressors $z$.

The terms $\delta(E)$ and $\delta(V)$ are the functional derivatives of each of the elements of $E$ and $V$, respectively. Consider a location shift $\varepsilon$ that only affects the distribution of years of education $h$. Then we can calculate the functional derivatives of each element in $E$ and $V$ as follows:

\begin{enumerate}[(i)]

\item \textit{First-order moments of $x$}

First, analyze the effect on $E_k$:

\begin{equation}
  \delta(E_k) = \lim_{\varepsilon \to 0} \frac{E[(h+\varepsilon)^k] - E(h^k) }{\varepsilon} = E\left[\lim_{\varepsilon \to 0} \frac{(h\varepsilon)^k - h^k }{\varepsilon}\right] = E(kh^{k-1}) = kE_{k-1}  \tag{a.3}\label{eq:a.3}
\end{equation}

In addition, the vector $E_z=E(z)$ does not change with a horizontal translation in $\varepsilon$, therefore it is evident that $\delta(E_z)=0 $.

Substituting all this in $\delta(E)$ gives the first part of (3.6).

\item \textit{Variance and covariance between $h^j$ and $h^k$}

The origin of this block of matrix $V$ is the inclusion of $h$ and $h^2$ as part of the covariates $x$. Note that $V_{jk}=E_{j+k}-E_jE_k$, then using the result (a.3) follows that:
\begin{align*}
\delta(V_{jk}) &= \delta(E_{j+k}-E_j E_k ) \\
			&=\delta(E_{j+k})-\delta(E_k )E_j - \delta(E_j ) E_k \\
			&= (j+k)E_{j+k-1} - kE_{k-1}E_{j} - jE_{j-1}E_{k}  \\
			&= kV_{j(k-1)} + jV_{(j-1)k}  \tag{a.4}\label{eq:a.4}
\end{align*}

para $j=0,1,2$ y $k=0,1,2$.

\item \textit{Covariance between $h^k$ and the regressors $z$}

Again, the inclusion of $h$ and $h^2$ together with the rest of the covariates $z$ results in this block of the matrix $V$. First, note that if $k=0$, the vector $M_{z0}$ is not affected by a location shift in $h$ variable, therefore $\delta(M_{0z})$ is a vector of zeros of dimension $Q$:
\begin{align*}
\delta(M_{0z}) &= 0_{1 \times Q}  \tag{a.5}\label{eq:a.5}
\end{align*}

For $ k> 1 $, each element is analyzed separately. Let $z_q$ be a covariable in $z$. The element $q$ of the vector $M_{kz}$ is $Cov(h^k, z_q) = E(h^k z_q)-E (h^k)E(z_q)$, therefore:
\begin{align*}
\delta[Cov(h^k,z_q )] 	&= \delta[E(h^k z_q )-E(h^k )E(z_q )]  \\
				 	&= \delta[E(h^k z_q )]-\delta[E(h^k )]E(z_q )  \\
				 	&= kE(h^{k-1} z_q ) - kE(h^{k-1})E(z_q )	\\
				 	&= kCov(h^{k-1},z_q)
\end{align*}

for $k=1,2$ and $q=1,2,…, Q$, where result (a.3) has been used together with the fact that $E(z_q)$ does not change before a translation of $h$. Moreover,
\begin{align*}
\delta[E(h^k z_q)]=  E\left[ \lim_{\varepsilon \to 0} \frac{(h+\varepsilon)^k - h^k}{\varepsilon} \cdot z_q \right]=kE(h^{k-1}z_q)
\end{align*}

Then, the vector $M_{kz}$ changes as follows:
\begin{align*}
\delta[M_{kz}] = kM_{(k-1)z} \tag{a.6}\label{eq:a.6}
\end{align*}

for $k=1,2$. Note that when $k=1$, the change in $M_{1z}$ is a null vector of dimension $Q$, because $M_{0z}$ is a vector of zeros.

\item \textit{Variances and covariances of $z$}

These moments do not depend on the $h$ distribution, therefore $\delta(M_{zz})$ is a null matrix of dimension $Q \times Q$:
\begin{align*}
\delta[M_{zz}] = 0_{Q \times Q} \tag{a.7}\label{eq:a.7}
\end{align*}

Substituting the results (a.4) - (a.7) in $\delta(V)$ gives as a result the second part of equation (3.6).

\end{enumerate}

\end{document}